\documentclass{article}
\usepackage{graphicx}
\usepackage{amsfonts}
\usepackage{amsmath}
\usepackage{amsthm}
\usepackage{authblk}
\usepackage{placeins}
\usepackage{hyperref}
\usepackage{algorithm}
\usepackage{algpseudocode}
\usepackage{amssymb}

\newtheorem{definition}{Definition}[section]
\newtheorem{assumption}[definition]{Assumption}

\newtheorem{prop}[definition]{Proposition}
\newtheorem{corollary}[definition]{Corollary}
\theoremstyle{remark}

\numberwithin{equation}{section}

\newcommand{\Apply}{\mathsf{Apply}}
\newcommand{\enc}{\mathsf{enc}}
\newcommand{\LH}{\mathsf{LH}}

\usepackage[
backend=biber,
style=alphabetic,
sorting=ynt
]{biblatex}
\title{Sei Giga\\
\small v2.0}
\author[]{Benjamin Marsh}
\author[]{Steven Landers}
\author[]{Jayendra Jog}
\author[]{Alejandro Ranchal-Pedrosa}
\affil{Sei Labs}
\date{June 2026}

\begin{document}

\maketitle

\begin{abstract}
Sei Giga is a multi-proposer EVM layer 1 architecture that separates ordering finality from execution and later state attestation. Consensus finalizes a deterministic order over transactions using the Autobahn consensus protocol, executors reconstruct the same canonical state transition off the consensus critical path, and validators later attest to compact divergence digests over the executed write log. The design goal of Giga is to allow for efficient, fast, and fair on-chain trading which requires very high throughput, low latency, and the ability to bound censorship and MEV risk. This document describes both the launch architecture and the near term extensions.
\end{abstract}

\section{Introduction}
In this work, we present Sei Giga, the first multi-proposer (sometimes called multiple concurrent proposer, multi-concurrent proposer, or MCP, in both this and other works) EVM layer-1 blockchain. Sei Giga uses Autobahn \cite{giridharan2025autobahnseamlesshighspeed} as a consensus protocol, which allows for over $5$ gigagas throughput \cite{5giga} and sub $250$ms finality under standard BFT-style security assumptions, which has already been achieved internally in a 40 node geo-distributed network across 20 regions. The MCP design of Giga means it is particularly well suited for trading applications as censorship resistance is provided by default, and several approaches to MEV mitigation are available to the end user. Sei Giga uses a new storage layer, and asynchronous execution divergence commitments over executed write sets rather than a hot path Merkle structure. As a result, Sei Giga is able to scale to allow for web2 style usage while ensuring quick finality in settings where it is necessary such as trading, on a verifiable and secure public ledger. Sei Giga is a decentralized, permissionless Proof-of-Stake EVM chain capable of running standard EVM smart contracts written in typical EVM languages such as Solidity and Vyper. Sei Giga's EVM support is equivalent to mainchain Ethereum with the exception of EIP-4844, PREVRANDAO, the state root, the block gas limit, and the transaction fee mechanism, and Sei Giga will continue to support EVM upgrades to maintain near parity. As is standard in Proof-of-Stake blockchains, Sei Giga is maintained by a set of staked nodes which act as validators and executors in the network responsible for ensuring consensus is met and blocks are run, the nodes are staked and eligible for block rewards to ensure an economic incentive to keep the chain running, and subject to slashing of the staked coins in the case of malicious behavior. Giga uses a three step pricing model for transactions, with a standard gas style execution fee priced according to a 1559 style demand function, an ordering fee where the priority fee becomes strictly enforced for ordering, and a distribution pricing fee to price duplicates.

In order to ensure speedy consensus, Sei Giga reaches initial consensus over the inclusion of transactions in a block and not the state of the chain given that block, the transactions then have a deterministic ordering across all blocks in the slot. Given the deterministic nature of EVM execution, this allows the chain to asynchronously execute blocks post-finality in order to reach an agreed state in a later block. The split block processing model allows Sei Giga to avoid execution bottlenecks by slowing consensus. Gas and other associated fees and rewards on the chain are handled by the underlying native coin of the chain, SEI. Sei Giga does not utilize a mempool, and transactions are immediately included by the node. Sei Giga supports multiple types of clients, including validator nodes, which participate in both consensus and execution; a full node, which consists of the RPC layer and the execution layer for the reading path; light nodes, which consist of the RPC layer alone, and data nodes, which exists purely to serve recent data. Sei Giga utilizes direct node-to-node networking preventing propagation across nodes for performance purposes. Transactions when submitted to the RPC are randomly allocated, in a stake weighted style, to a validator reducing the impact of validator targeted spam attacks, transactions may be submitted to multiple validators for censorship resistance leading to a fairer transaction inclusion process. A partial refund of the transaction tip will be refunded if multiple copies of the transaction exist and only a single copy is executed.

\begin{definition}[Ordering finality]
Let $B_n$ denote the block induced by the $n$th committed cut. We say that $B_n$ has \emph{ordering finality} once the consensus layer finalizes the unique total order of transactions induced by that cut.
\end{definition}

That is to say that consensus is initially reached over the set of transactions in the cut in all blocks. The sequence can then be deterministically derived from the set.

\begin{definition}[State attestation finality]
Let $D_n$ denote the divergence digest for the execution of $B_n$. We say that $B_n$ has \emph{state attestation finality} once a later finalized block includes a valid $\frac{2}{3}$ signed attestation to $D_n$.
\end{definition}

Unless stated otherwise, latency claims in this document refer to ordering finality rather than state attestation finality.

\begin{figure}[hbt!]
    \centering
    \includegraphics[width=1\linewidth]{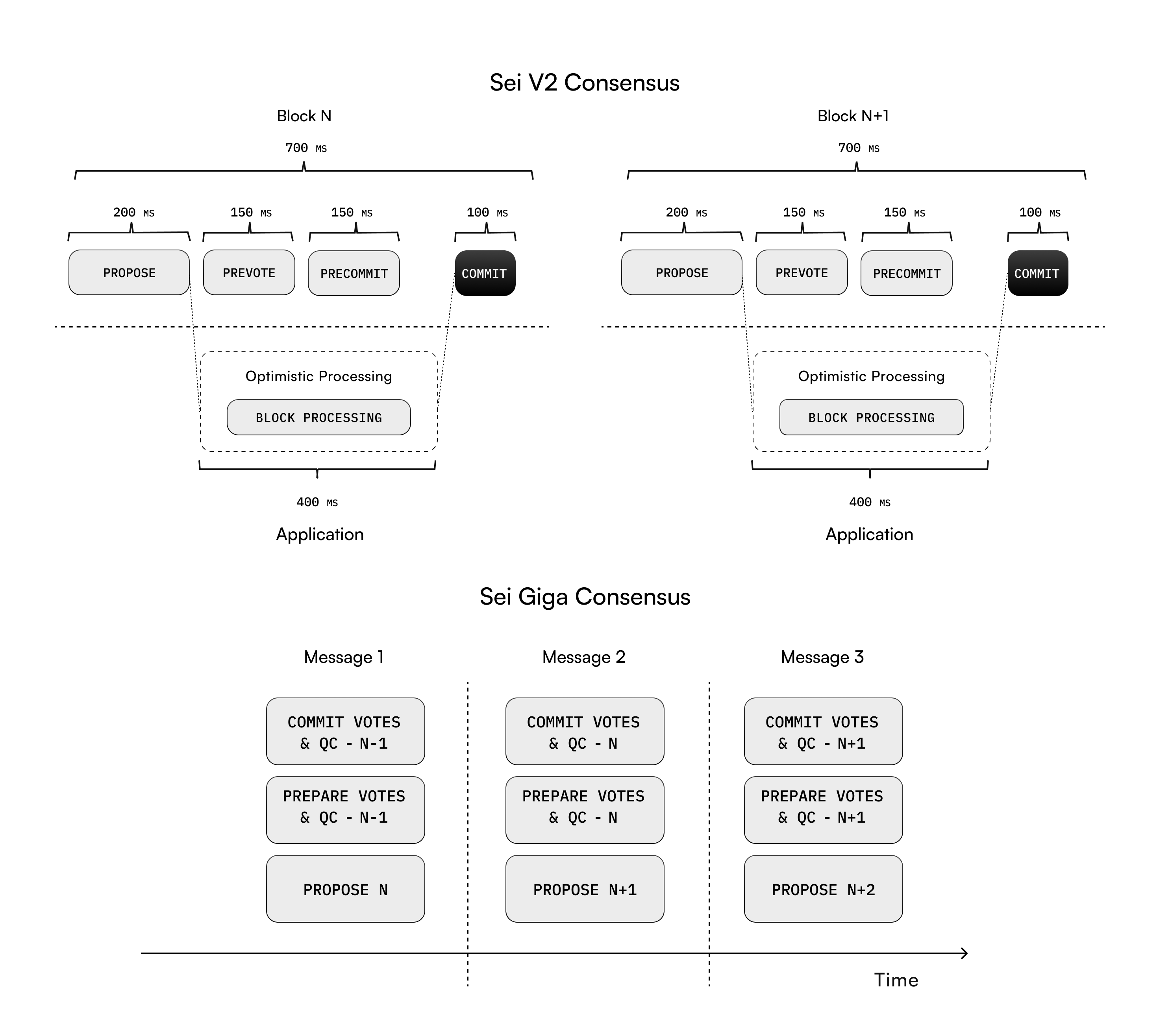}
    \caption{v2 vs Giga}
    \label{fig:v2-vs-giga}
\end{figure}
\subsection{Comparison to Sei v2}
Autobahn achieves over $50$× higher throughput than Tendermint by fundamentally rethinking how consensus and data availability are managed in a distributed system. In Tendermint, each block is processed sequentially where a single leader proposes a block, validators must download and fully execute the block transactions, and then consensus is reached in multiple rounds before the next block is proposed. This tight coupling between data dissemination, execution, and consensus means that every block decision incurs significant overhead and waiting time, resulting in a throughput limited by the speed of one proposer and the time taken to complete all three rounds of communication.

Autobahn, on the other hand, decouples these processes to dramatically increase performance. Instead of relying on one leader at a time, every node continuously disseminates its own stream of data proposals in independent lanes. This multi-proposer architecture eliminates the bottleneck of a single proposer by allowing parallel data proposals. In this system, each node operates as its own proposer, continuously creating a chain of proposals. The consensus layer periodically commits a “tip cut,” which is a compact snapshot that aggregates the latest proposals (tips) from every lane. Since each tip implicitly references the entire history of its lane, this mechanism allows multiple blocks to be committed in one consensus instance without having to process each block individually.
A critical innovation in Autobahn is its decoupling of data availability from consensus. In Tendermint, validators must download and verify the entire block before they can cast their votes, which ties the consensus process directly to the full dissemination of data. Autobahn, however, uses proofs of availability (PoA) to certify that a block’s data is accessible without requiring every validator to download it immediately. Validators can vote based on these compact proofs while data synchronization occurs asynchronously in the background. This separation means that the ordering of transactions is not slowed down by the need to fetch full block data during the consensus round.

Autobahn further reduces latency by lowering the number of communication rounds. While Tendermint typically requires three full rounds of message exchanges to finalize a block, Autobahn’s optimized protocol reaches consensus in $1.5$ roundtrips in Sei Giga. Fewer rounds of communication directly translate into lower per-block latency and, consequently, higher overall throughput.
Additionally, Autobahn supports parallel slot execution where new consensus slots can begin before previous ones are fully committed. This pipelining ensures that the system continuously processes proposals without waiting for each individual block’s final commitment. Combined with an efficient view-change mechanism that quickly recovers from network blips or faulty leaders, Autobahn maintains high performance even under suboptimal conditions. Alongside pipelined block production, Sei Giga’s Autobahn allows for asynchronous block execution, unlike Tendermint, meaning we reach consensus over the ordering of transactions in a block and agree on the state later; this ensures that execution never becomes a bottleneck in the block production process.

The execution process is fully pipelined with parsing, address recovery, and signature verification happening in parallel to avoid bottlenecks in the execution process. Post-processing of the block, such as receipt generation, is handled out of the hot path meaning the next block can start execution while the previous block is undergoing post-processing.
By default, the frequently accessed state of Sei Giga is stored in RAM for optimum performance, with the full state stored on disk. The system is designed to minimize disk reads by ensuring most reads for a transaction (such as balance checks) are in memory, while disk writes are performed asynchronously solely for recovery purposes.

Overall, by decoupling execution from consensus, enabling multi-proposer parallelism, committing multiple blocks in a single tip cut, reducing the number of communication rounds, and separating data availability from the critical consensus path, the Sei Giga upgrade brings significant improvements over Sei v2. The faster consensus algorithm reduces voting rounds from $3$ to $1.5$, resulting in a $2$x improvement in consensus efficiency. The novel data availability layer with its multi-producer design delivers a $70$x improvement in block production with $180$ blocks versus $2.5$ previously while still ensuring strong Byzantine Fault Tolerance.

\section{Async Execution}
Sei Giga operates on the basis that once ordering is agreed upon in a block that the deterministic execution of that block will always result in the same output state, that is to say that consensus can be reached over just the contents and ordering of a block without execution.

\begin{definition}[Block context]
For finalized block $B_n$, let $c_n$ denote the full block context supplied to the EVM, including every block scoped field that may affect execution.
\end{definition}

\begin{definition}[Canonical block transition]
Fix a pre-state $S_{n-1}$, a block context $c_n$, and an ordered transaction sequence $B_n=(\mathrm{tx}_1,\ldots,\mathrm{tx}_m)$. Define
\[
\Apply(S_{n-1},c_n,B_n)=(S_n,R_n,G_n,\Delta_n)
\]
to be the canonical execution of $B_n$ from $S_{n-1}$ under $c_n$, returning the post-state $S_n$, receipts $R_n$, gas accounting $G_n$, and canonical write log $\Delta_n$.
\end{definition}

\begin{prop}[Deterministic block transition]\label{prop:det-exec}
Fix $S_{n-1}$, $c_n$, and $B_n$. Assume every honest executor uses the same execution semantics and that any parallel implementation is serializable to the block order of $B_n$. Then $\Apply(S_{n-1},c_n,B_n)$ is unique.
\end{prop}

\begin{proof}
The EVM specification defines a sequential state transition function, given a fixed pre-state, block context, and transaction sequence, each transaction's output is uniquely determined by the state left by its predecessor. By assumption, every honest executor uses the same semantics, and any parallel implementation is serializable to block order. Hence the composite transition $\Apply(S_{n-1},c_n,B_n)$ is unique.
\end{proof}

\begin{figure}[hbt!]
    \centering
    \includegraphics[width=1\linewidth]{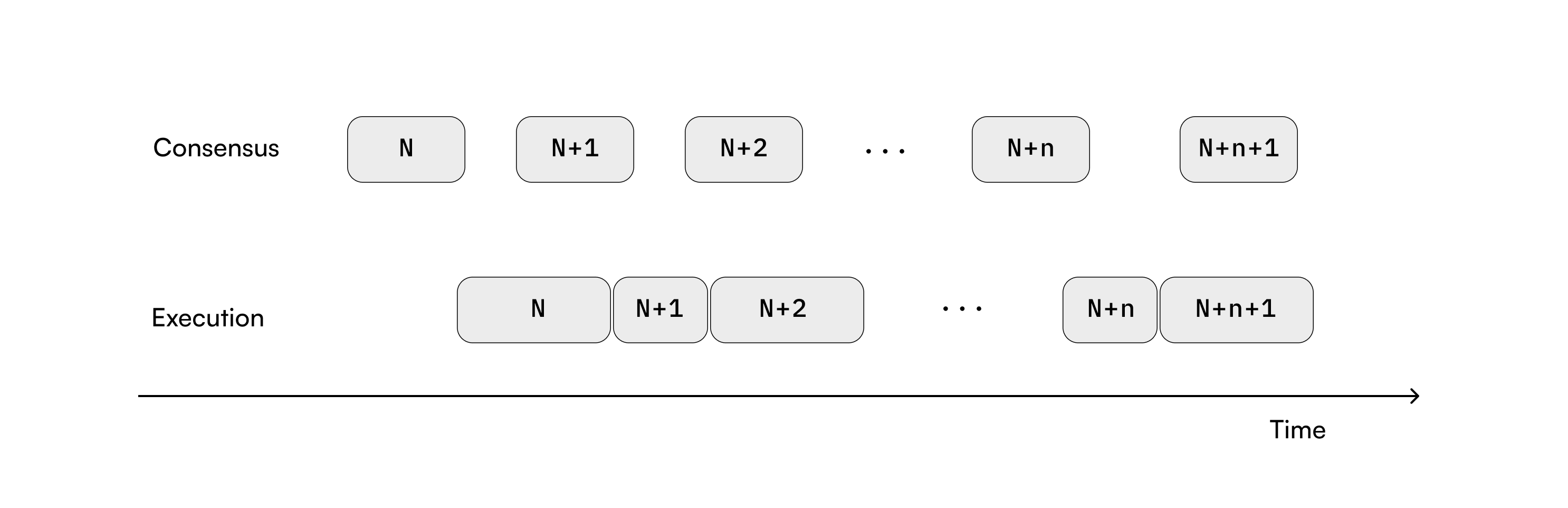}
    \caption{Async block execution}
    \label{fig:para2}
\end{figure}

\subsection{State consensus}
Given that execution itself is not needed for ordering consensus, Sei Giga can keep execution off the consensus critical path and run it asynchronously after block finality. Once a block $n$ finalizes, each executor deterministically computes the post-state transition together with a canonical write log, receipts, gas accounting, and the ordered set of touched keys. Validators then gossip a compact divergence digest for that executed transition and later include a $\frac{2}{3}$-signed attestation to the digest in a subsequent block $n+x$, for some $x \in \mathbb{Z}$ with a bound based on execution timing. During this period the chain continues to produce blocks $n+1, n+2, \dots$. This delayed attestation allows validators to confirm that they have not deviated from the agreed execution path without forcing full state agreement into the hot path. If a small minority of nodes diverges due to software or hardware faults, the chain can continue while the fault is localized. If divergence exceeds the standard Byzantine threshold, the chain pauses as in any other BFT system. A transaction that reverts remains reverted, but this does not invalidate the execution of the rest of the block. The production path and the execution attestation path are therefore separated by design.

\begin{prop}[Uniqueness of an attested divergence digest]
\label{prop:digest-unique}
Assume $n=3f+1$ validators, at most $f$ Byzantine faults, and that each honest validator signs at most one divergence digest for any fixed finalized block. Then two distinct digests $D_n \neq D'_n$ cannot both obtain valid $\frac{2}{3}$ signed attestations for block $n$.
\end{prop}

\begin{proof}
Any two signer sets of size at least $2f+1$ intersect in at least
\[
(2f+1)+(2f+1)-(3f+1)=f+1
\]
validators. Since at most $f$ validators are Byzantine, the intersection contains at least one honest validator. That validator would have signed both $D_n$ and $D'_n$ for the same block, contradicting the signing rule.
\end{proof}

Accordingly, conflicting attestations for the same finalized block imply slashable equivocation or a failure of the honest signing assumption.

\subsection{Execution client}
The client only handles tx processing and does not handle any other functionality such as tracing or log searching. Blocks are received and processed in parallel meaning only the block execution step happens sequentially, that is to say that all pre-processing and checks such as signature recovery happen in parallel, though only a single block will be executed at once. Post-processing of the block, such as receipt generation, is handled out of the hot path meaning the next block can start execution while the previous block is undergoing post-processing. The client is optimised to reduce disk reads by ensuring \textit{most} reads for a transaction (check balance) are in memory. Disk writes are all asynchronous.
\subsection{Pipelining and encoding}
The execution process is fully pipelined with parsing, address recovery, and signature verification happening in parallel to avoid bottlenecks in the execution process by ensuring these otherwise potentially blocking processes.

A flat encoding format is also utilized to ensure cheaper transaction decoding. The transaction encoding is a flat, length-prefixed layout. Each field, such as nonce, gas limit, and signature, is written out in a known order. Variable-length fields are prefixed with a small byte indicating their length. This design eliminates nested structures, enabling fast, single-pass decoding, straightforward zero-copy parsing and overall lower overhead. Once all fields are consumed, any remaining bytes are assumed to be the transaction input. This approach ensures minimal serialization overhead.
\begin{algorithm}[H]
\caption{Parsing a Transaction Payload}
\label{alg:parse-payload-simplified}
\begin{algorithmic}[1]
\Require \textit{payload} (\texttt{[]byte}) 
\Ensure \textit{tx} is populated

\State $ptr \gets 0$
\State Read and store \(\mathit{Type}\), \(\mathit{ChainID}\), \(\mathit{Sender}\), \(\mathit{To}\), 
  \(\mathit{Value}\), and other fields by consuming the appropriate number of bytes from \(payload\).
\If{next byte indicates a contract creation}
   \State \(\mathit{To} \gets \text{nil}\)
\EndIf
\State Read an integer \(n\) for the number of access-list entries
\For{$i = 1$ to $n$}
    \State Read an integer \(m\) for the number of storage keys
    \For{$j = 1$ to $m$}
        \State Read each storage key from \(payload\)
    \EndFor
    \State Add the address and its keys to \(tx.\mathit{AccessList}\)
\EndFor
\State Remaining bytes become \(\mathit{Input}\)
\end{algorithmic}
\end{algorithm}
\section{Consensus and Data Dissemination}

Autobahn \cite{giridharan2025autobahnseamlesshighspeed} is a BFT style consensus protocol, used as a Proof-of-Stake mechanism in Giga, designed to be performant in the partial synchrony model like HotStuff \cite{yin2019hotstuffbftconsensuslens}. Where other highly performant protocols offer low latency in fault free synchronous periods or robust recovery from interrupts. To bridge the gap between traditional view based protocols which suffer during blips and DAG based protocols which offer non-optimal latency during good intervals, Autobahn offers a hybrid approach with a parallel asynchronous data dissemination layer with a low latency, partially synchronous consensus mechanism. Thus Autobahn is able to avoid the hangovers of traditional protocols while matching BFT throughput with half the latency. A stake weighted leader selection process akin to Tendermint is used to select leaders. Formal discussions of the security of Autobahn can be found in \cite{giridharan2025autobahnseamlesshighspeed}.

\subsection{Overview and High-Level Architecture}
Autobahn separates data availability from ordering. At its core is a data dissemination layer that organizes proposals into independent lanes, each enhanced with a Proof of Availability (PoA). This ensures that data is available with minimal latency overhead even before being ordered into the global chain. The consensus layer then periodically commits a snapshot (a \textit{cut}) by ordering the latest certified tips from all lanes. 
\begin{figure}[hbt!]
    \centering
    \includegraphics[width=1\linewidth]{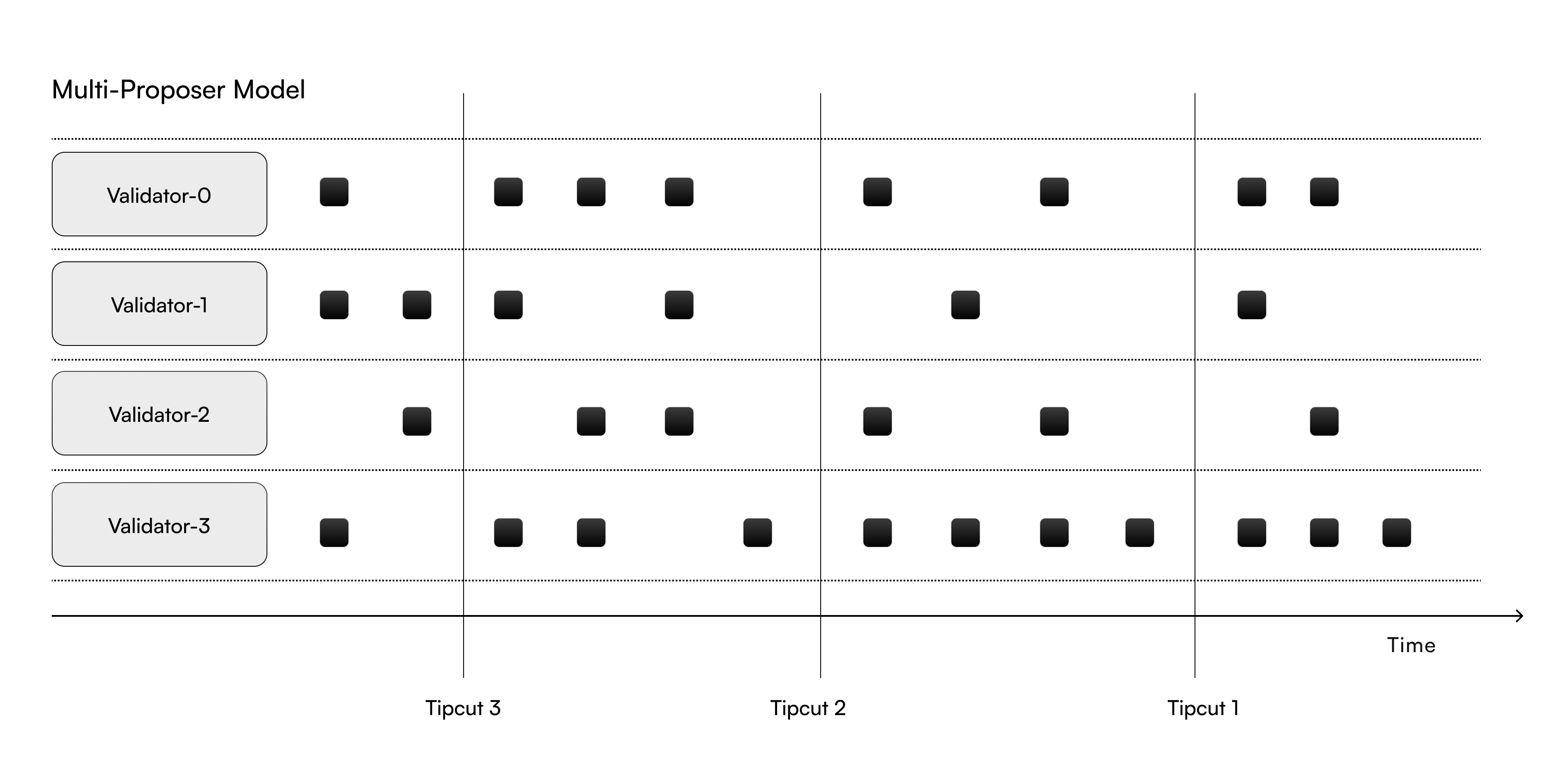}
    \caption{Multi-producer}
    \label{fig:multi-producer}
\end{figure}
\subsection{Data Dissemination Layer (Lanes) and PoA}
The DA layer’s sole job is to guarantee that every batch committed in a cut can later be downloaded by any honest node. We achieve this with an f+1 Proof‑of‑Availability certificate (PoA).
\paragraph{Lane Structure and Proposal Process.}
Each replica $r$ maintains its own lane $\ell_{r}$, in which it sequentially proposes new batches (or \textit{cars}) of transactions:
\[
\ell_{r}^{0}, \ell_{r}^{1}, \ell_{r}^{2}, \ldots
\]
When replica $r$ has a new batch of transactions $B$, it constructs a data proposal:
\[
\mathrm{Prop} \;=\; \langle \mathrm{pos}, B, \mathrm{parentRef} \rangle_{r},
\]
where $\mathrm{pos}$ denotes the sequence number in the lane and $\mathrm{parentRef}$ references the hash of the previous proposal in $\ell_{r}$.
\begin{figure}[hbt!]
    \centering
    \includegraphics[width=1\linewidth]{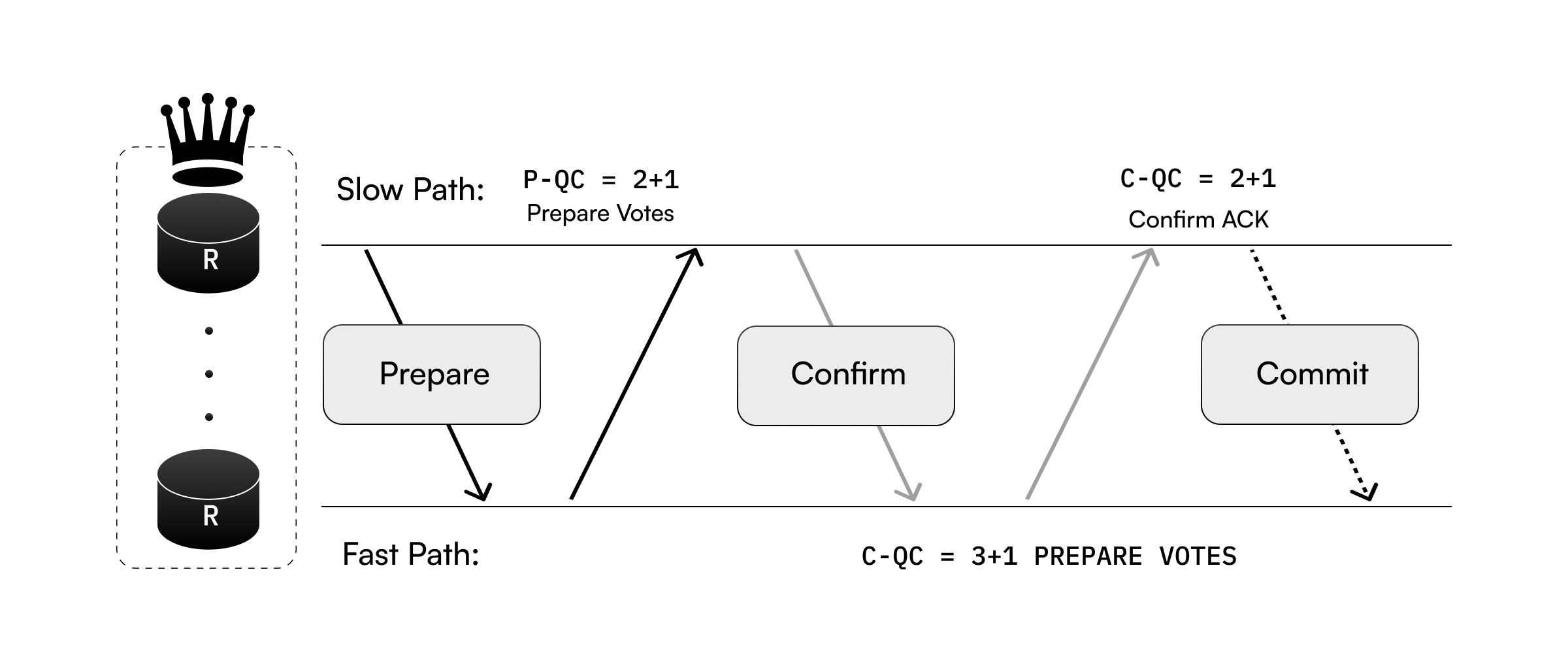}
    \caption{Prepare}
    \label{fig:prepare}
\end{figure}
\paragraph{Voting and PoA Certification.}
Each replica that receives and validates $\mathrm{Prop}$ (i.e., by checking that $\mathrm{parentRef}$ matches the previously voted proposal) returns a vote:
\[
\textit{Vote} = \langle \mathrm{digest}(\mathrm{Prop}) \rangle.
\]
Once $r$ collects $f+1$ matching votes, it assembles a PoA:
\[
\mathrm{PoA} \;=\; \bigl(\mathrm{digest}(\mathrm{Prop}), \{\sigma_i\}_{i \in Q}\bigr) \quad \text{with} \quad |Q| = f+1.
\]

\begin{prop}[Minimal availability guarantee of a PoA]
\label{prop:poa-availability}
If a proposal $\mathrm{Prop}$ obtains a valid PoA with $f+1$ votes in a system with $n=3f+1$, then at least one correct replica possessed the full proposal data at the time it voted.
\end{prop}

\begin{proof}
A quorum of size $f+1$ contains at least one correct replica because at most $f$ replicas are Byzantine. Correct replicas vote only after receiving and validating the full proposal data. Hence at least one correct replica possessed the full proposal when the PoA was formed.
\end{proof}

The PoA is the minimal availability guarantee needed to justify off-path retrieval. It is not, by itself, a long term retention guarantee, and no such guarantee follows from quorum intersection, a $2f{+}1$ commit quorum and an $f{+}1$ PoA quorum are only guaranteed to intersect in a single replica, which may be faulty. Long term availability instead follows from the retrieval path itself.

\begin{assumption}[Vote retention]\label{ass:vote-retention}
A correct replica that votes for a proposal stores the full proposal data and serves it on request at least until every correct replica has retrieved it.
\end{assumption}

\begin{corollary}[Retrievability of committed data]\label{cor:committed-retrievable}
Under Assumption~\ref{ass:vote-retention}, every batch referenced, directly or transitively, by a committed cut is eventually held by every correct replica.
\end{corollary}

\begin{proof}
A correct replica votes for a proposal at position $\mathrm{pos}$ in a lane only after validating that its $\mathrm{parentRef}$ matches the proposal it previously voted for, so a correct member of the tip's PoA quorum has received every proposal in that lane's prefix. By Proposition~\ref{prop:poa-availability}, the tip's PoA quorum contains at least one such correct replica. Every correct replica must execute committed cuts and therefore requests any batches it is missing off the critical path, which the correct holder serves by Assumption~\ref{ass:vote-retention}. Hence each referenced batch propagates to all correct replicas, after which durability is a local storage property rather than a quorum property.
\end{proof}

We note that strengthening the PoA quorum to $2f{+}1$ would guarantee $f{+}1$ correct holders at certification time, at the cost of certification latency. Giga retains the $f{+}1$ quorum because commitment itself triggers full replication among correct replicas via the execution path, so the larger quorum buys durability only in the short window between certification and commit.

\begin{figure}[hbt!]
    \centering
    \includegraphics[width=1\linewidth]{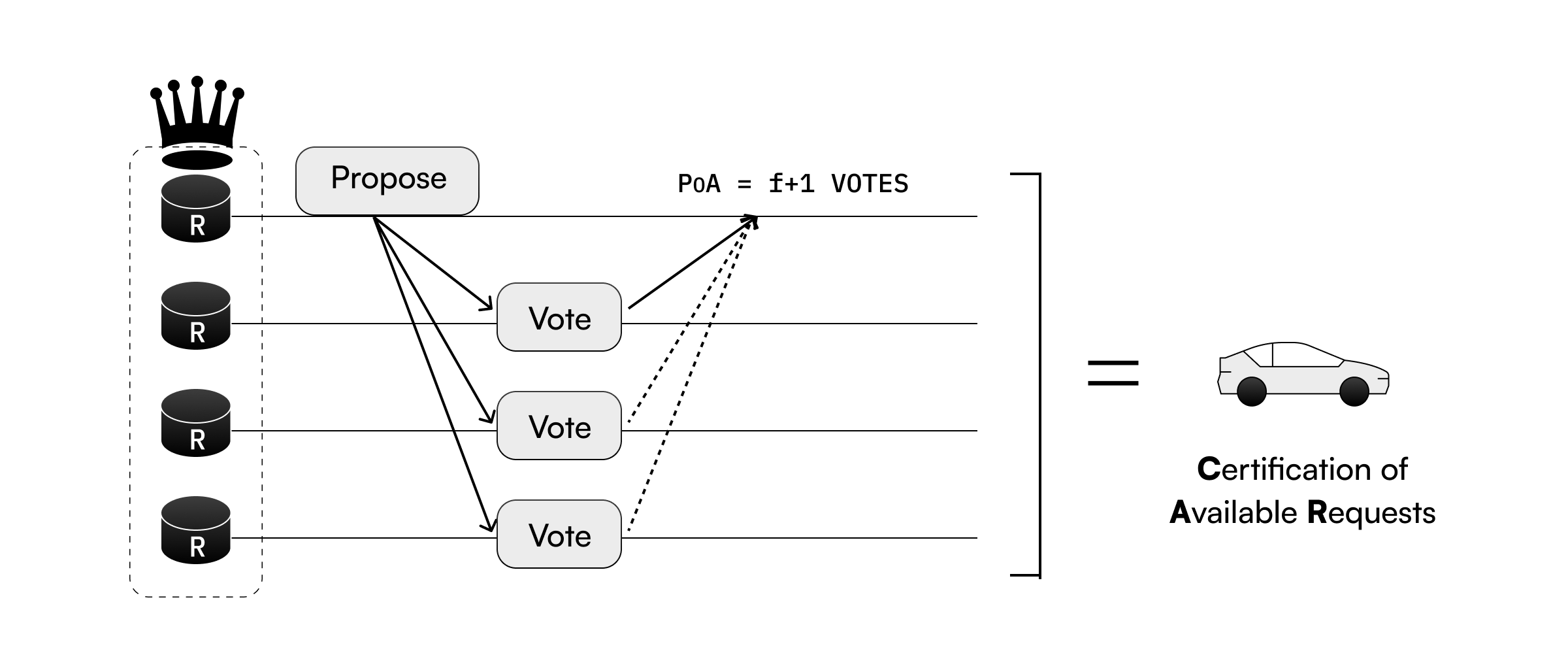}
    \caption{Proposal}
    \label{fig:proposal}
\end{figure}

\paragraph{Instant Referencing.}
Because proposals in lane $\ell_{r}$ are chained, referencing the tip (i.e., the most recent proposal with a PoA) implicitly attests that all previous proposals in that lane are available. This transitive guarantee reduces synchronization overhead when lanes are later merged in the consensus process. 
\begin{figure}[hbt!]
    \centering
    \includegraphics[width=1\linewidth]{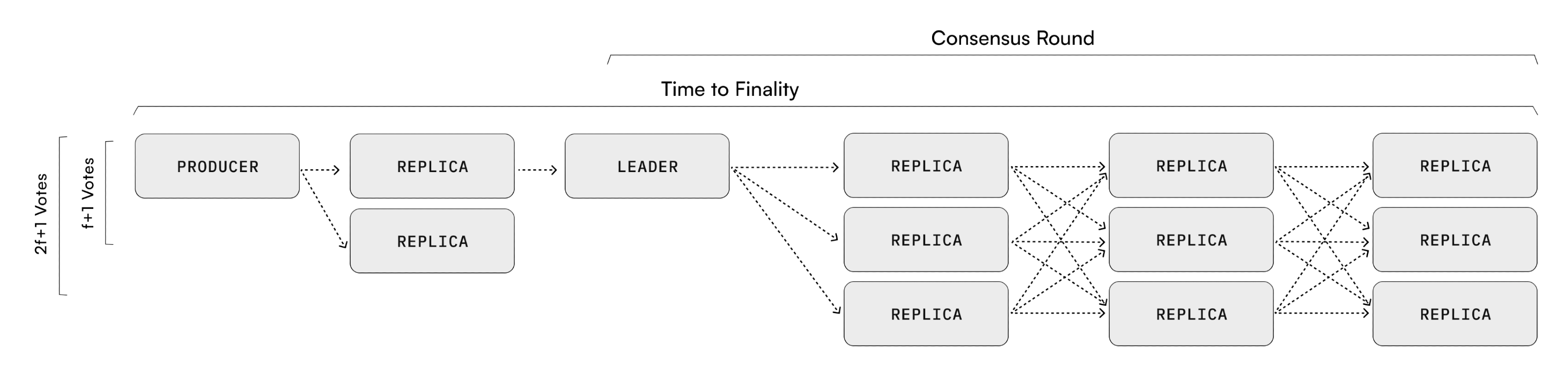}
    \caption{DA and consensus stages with vote reqs}
    \label{fig:da-stages}
\end{figure}
\subsection{Data Availability and Final Execution}

\paragraph{Data Synchronization (DA) and Asynchronous Retrieval.}
Since each lane’s tip carries a PoA, any replica missing a batch can retrieve it asynchronously from a replica within the PoA set. When a global $\mathrm{Cut}$ is committed, any missing batches must be fetched off the critical path. If the leader is correct and the network is synchronous, Autobahn commits a new proposal in at most two communication rounds per slot. When a leader fails to make progress, replicas revert to a standard view change mechanism and elect a new leader via a timeout certificate.

\paragraph{Final Global Ordering.}
Let $M$ denote the deterministic merge rule that maps a committed cut $C$ to a unique ordered transaction sequence $B(C)=M(C)$. Every correct executor applies the same rule $M$ to the same committed cut.

\begin{corollary}[Consistent execution from a committed cut]
\label{cor:cut-to-exec}
If all correct executors observe the same committed cut $C$, then they derive the same ordered transaction sequence $B(C)$ and therefore compute the same canonical block transition for $B(C)$.
\end{corollary}

\begin{proof}
Consensus gives all correct executors the same committed cut $C$. Because $M$ is deterministic, all correct executors derive the same ordered transaction sequence $B(C)=M(C)$. Proposition~\ref{prop:det-exec} then implies a unique canonical block transition.
\end{proof}

\subsection{Consensus Layer (Cut of Tips)}

\paragraph{Global Ordering via Cuts.}
The consensus layer periodically commits a snapshot of the system by collecting the latest certified tips from all lanes:
\[
\mathrm{Cut} \;=\; \{\ell_{1}[\mathrm{tip}],\;\ell_{2}[\mathrm{tip}],\ldots,\ell_{n}[\mathrm{tip}]\}.
\]
A designated leader (selected via a stake weighted process similar to Tendermint) initiates the ordering process.
\begin{figure}[hbt!]
    \centering
    \includegraphics[width=1\linewidth]{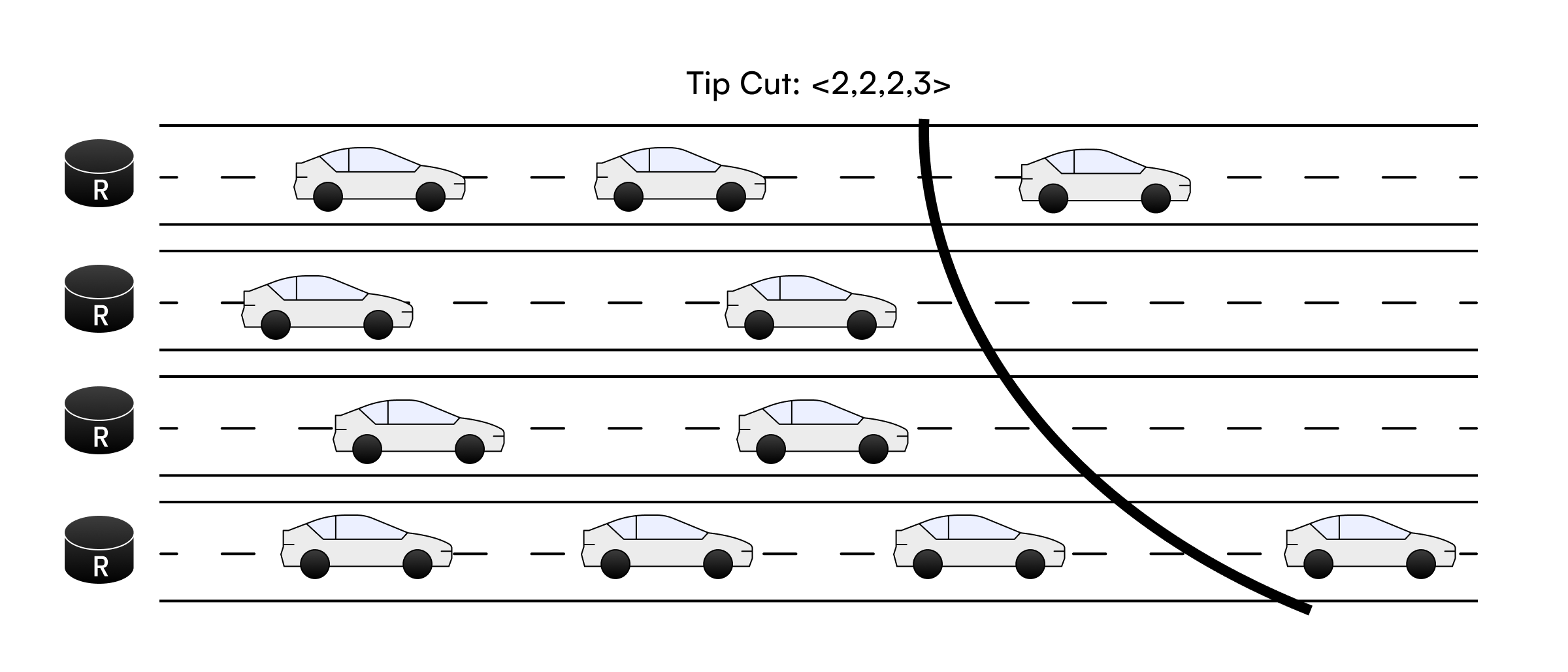}
    \caption{Tip cut}
    \label{fig:tip-cut}
\end{figure}
\paragraph{Protocol Phases and Pipelining.}
Autobahn implements a two phase BFT agreement protocol:
\begin{enumerate}
    \item \textbf{Prepare:} The leader collects the most recent certified tips from all lanes and bundles them into a proposal. Honest replicas then broadcast their prepare votes.
    \item \textbf{Commit:} Once enough votes are collected, a CommitQC is formed, finalizing the blocks.
    \item \textbf{Confirm:} If the leader gathers only $(n-f)$ votes, it enters a confirm phase and waits for additional acknowledgments until a commit certificate is achieved with $2f+1$ confirm messages.
\end{enumerate}

\paragraph{Pipelining Benefit.} Through the use of quadratic communication and pipelining, Sei Giga achieves a slow path round trip of 1.5 rather than 2.5. That is to say that the leader sends a proposal for tip cut $N$, triggering nodes to send their prepare votes. Once enough votes are collected, they form a PrepareQC. After that, nodes send their commit votes, and once enough are collected, a CommitQC is formed, finalizing the blocks in $N$. While $N$ is in its commit phase, the leader for tip cut $N+1$ may start sending its proposal. In other words, even though each block goes through two distinct rounds, pipelining allows the next block’s proposal to start during the commit phase of the previous block, effectively reducing the overall latency to 1.5 rounds.

\paragraph{Parallel Slots.}
To minimize delay between successive blocks, consensus slots are pipelined. Once a replica sees the Prepare message for slot $s$, it can begin slot $s+1$ without waiting for slot $s$ to fully commit, ensuring that proposals arriving just too late for one slot do not cause extra round delays.

\subsection{System Model, Security, and Reliable Inclusion}

\paragraph{System Model and Assumptions.}
\begin{itemize}
    \item \textbf{Replicas:} There are $n=3f+1$ replicas; up to $f$ may be faulty.
    \item \textbf{Authentication:} All messages are transmitted over authenticated, point-to-point channels and cryptographic primitives are unforgeable.
    \item \textbf{Partial Synchrony:} Safety is unconditional; liveness requires eventual network stabilization such that message delays respect known upper bounds.
    \item \textbf{Clients:} There is no upper bound on potentially faulty clients; clients submit transactions that are batched into proposals.
\end{itemize}

\paragraph{Security and Reliable Inclusion.}\label{sec:autobahn-security}
Autobahn’s design ensures that no two conflicting proposals can both obtain a valid commit certificate in the same slot. Key points include:
\begin{enumerate}
    \item Once a proposal receives $f+1$ votes, at least one correct replica stores the data.
    \item A correct leader must include such proposals in a future $\mathrm{Cut}$.
    \item Faulty proposers cannot indefinitely disseminate data without eventual inclusion.
\end{enumerate}
This mechanism prevents Byzantine replicas from causing wasted dissemination and protects against censorship beyond a bounded delay.

\begin{figure}[hbt!]
    \centering
    \includegraphics[width=1\linewidth]{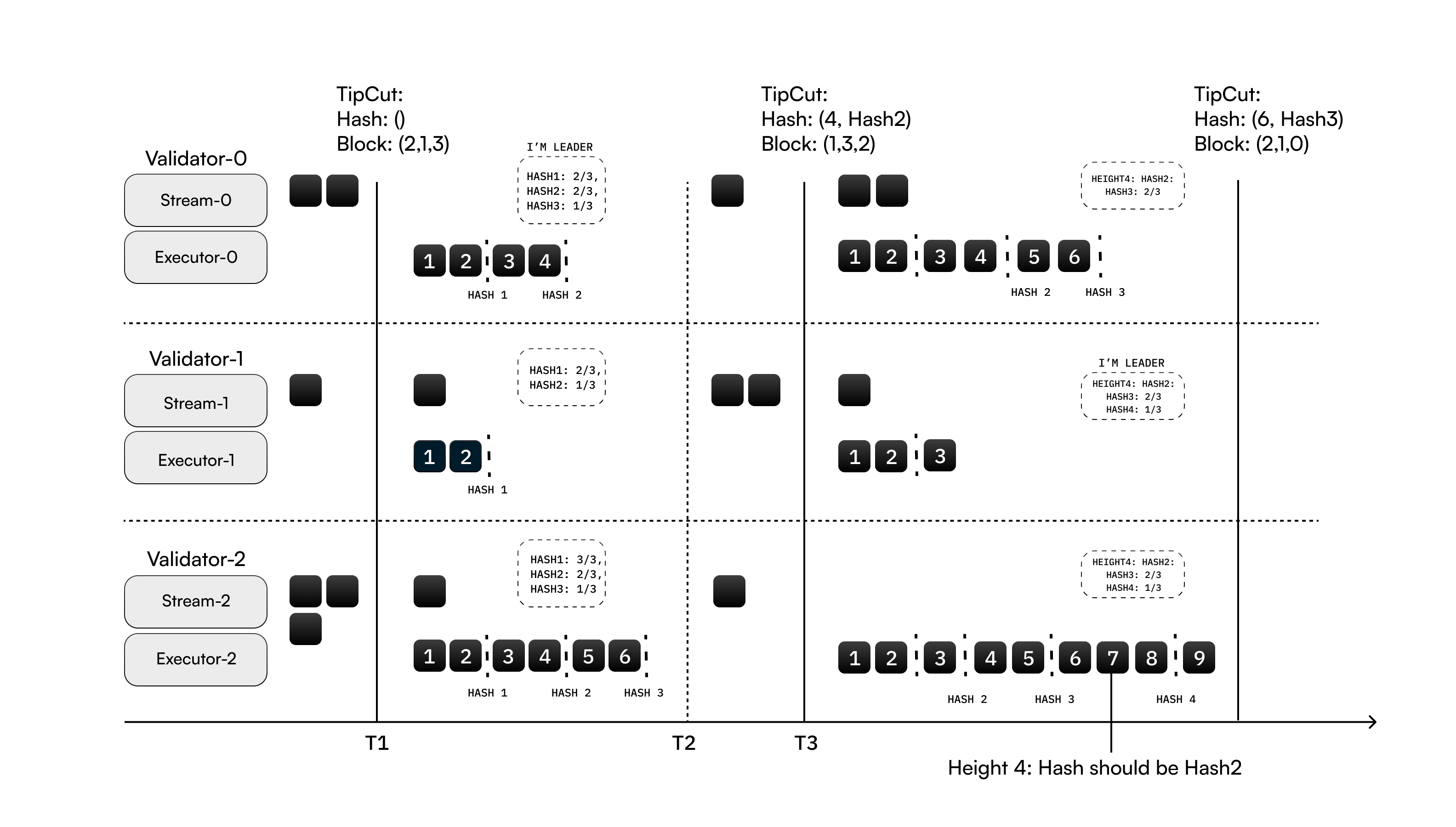}
    \caption{Async block execution}
    \label{fig:async-exec}
\end{figure}
\FloatBarrier

\subsection{Ambulance and backup leaders}\label{sec:ambulance}
Autobahn is intentionally single leader at the tip cut layer. A single leader proposes the cut, validators vote, and the slot commits without unnecessary coordination. The weakness is in the tail, if the designated cut leader is fail slow because of I/O contention, garbage collection, data synchronization, regional routing problems, or ordinary network blips, finality can inflate until timeout driven recovery starts. Aggressive timeouts create unnecessary view changes, conservative timeouts leave the protocol idle while latency spikes. Ambulance \cite{giridharan2026ambulance} addresses this problem by replacing a race against a timeout with a protocol rigged race between replicas. In Ambulance, the designated leader still has the structurally fastest path and should win in normal conditions. Non-leader replicas, however, do not wait idle for a timeout or a hedging delay. They execute useful protocol steps in their own lanes while the leader is trying to commit. The race is rigged in the leader's favour by giving the leader fewer protocol steps, so the protocol preserves the low latency leader path when the leader is healthy. If the leader is slow, the work already performed by other replicas feeds directly into a recovery path rather than being thrown away. This is the key distinction from both timeout triggered view changes and time based hedging and slowdown detection itself is productive. For Giga, the intended interpretation is to preserve Autobahn as the optimistic path while adding an Ambulance style recovery path for slow cut leaders. A single leader would still propose the tip cut and should still commit first in the median case. In parallel, a bounded backup set, or eventually the whole validator set, can assemble candidate cuts from the same certified lane tips and perform useful non-equivocation and persistence work. If the leader's certificate forms first, the slot commits exactly as today. If the leader lags, the protocol can finish from already existing recovery state rather than paying a full timeout penalty. This preserves Autobahn's data dissemination layer and pipelining strategy, Ambulance is explicitly compatible with Autobahn style motorization, where the consensus protocol uses Autobahn's data layer for high throughput. In essence we retain Autobahn performance when all is well and handle the edge cases where performance would have degraded in a more performant manner. Ambulance will form a core part of a future upgrade.

\section{Transaction Dissemination and MEV in Multiple Concurrent Proposer Chains}
Sei Giga's multi-proposer architecture changes the structure of MEV. In a sequential chain, the dominant question is usually what a single proposer, builder, or sequencer does with a transaction once it controls the block. In a multiple concurrent proposer chain, multiple blocks can become data available before their final execution order is fixed. The consequence is that MEV moves outward from a single private block-building step toward same tick inter-block races over who sees an order first, who can duplicate it, and how quickly it becomes actionable \cite{landers2025mcpmev}. The immediate user problem is dissemination. If a sender transmits the full transaction to many proposers, it improves censorship resistance and often lowers inclusion latency, but pays for this with replicated bandwidth and broader pre-inclusion information leakage. If instead the sender transmits to only a few proposers, system goodput improves, but the sender accepts weaker latency and censorship guarantees. This is not a cosmetic trade off. In latency sensitive settings, a short delay can destroy the economic value of the transaction entirely. For an MCP chain, dissemination policy is therefore part of the protocol's economics, not merely a networking detail \cite{ranchal2025sedna}. Recent work identifies several MEV channels that are specific to MCP execution \cite{landers2025mcpmev}. The first is the same tick duplicate steal, once a transaction is visible on the data dissemination layer, another proposer can copy it into its own lane and attempt to win the deterministic merge order. The second is proposer-to-proposer orderflow trade, where a proposer or relay that learns of an opportunity sells that information to a better positioned proposer. The third is the timing race created by proof-of-availability latency itself such that even when merge rules are deterministic, faster visibility changes who can react before execution. More broadly, MCP turns user multi-submission into a protocol level externality. If the chain leaves duplicate handling or merge policy underspecified, users seeking censorship resistance mechanically create new proposer side extraction opportunities \cite{landers2025mcpmev}. Some of these channels should be constrained directly by protocol rules. In particular, duplicate aware tip handling and deterministic priority respecting merge rules reduce the gain from same tick copying and narrow the space for discretionary extraction after data is visible \cite{landers2025mcpmev}. But this does not solve the dissemination problem itself. A user may still be forced to choose between broad transparent replication, which leaks information, and narrow submission, which weakens latency and censorship guarantees. A natural response to transparent order flow is to reach for an encrypted mempool. That is not the cleanest fit for Sei Giga. Giga does not rely on a global mempool and does not benefit from adding decryption coordination or new hot path cryptography merely to recreate one. More importantly, encrypting a transaction does not by itself solve the MCP dissemination problem where a user may still need to replicate the payload widely to get good latency and censorship resistance, thereby wasting bandwidth even if the payload is hidden.

\subsection{Merge rule: sort by tip}\label{sec:merge-rule}
The deterministic merge rule $M$ used by Sei Giga orders transactions from a committed cut as follows.

\begin{definition}[Tip priority merge rule]
Let $C = \{\ell_1[\mathrm{tip}], \ldots, \ell_n[\mathrm{tip}]\}$ be a
committed cut. For each lane $\ell_r$, let $\mathrm{txs}(\ell_r) = (t_{r,1}, \ldots, t_{r,k_r})$ denote the transactions newly committed by the tip of lane $r$ (i.e.\ those not included in a previous cut), ordered by their intra-lane position. The merge rule $M(C)$ is the sequence obtained by:
\begin{enumerate}
    \item Sorting the lanes by descending maximum transaction tip
      offered by any transaction in $\mathrm{txs}(\ell_r)$, breaking
      ties by replica index.
    \item Within each lane, preserving the intra-lane position order.
    \item Deduplicating: if the same transaction hash appears in
      multiple lanes, only the first occurrence in the merged sequence
      is retained; subsequent duplicates are dropped.
\end{enumerate}
\end{definition}

This rule is fully determined by the committed cut and requires no information beyond the finalized lane contents, ensuring every correct executor derives the same ordered transaction sequence from the same cut. The lane level sort by maximum tip gives proposers who carry higher value orderflow priority in the global ordering, creating a
direct incentive to include valuable transactions rather than withhold them. The duplicate drop rule ensures that users who submit to multiple lanes for censorship resistance do not pay execution costs more than once \cite{landers2025mcpmev}.

\begin{prop}[Determinism of the tip priority merge]
The merge rule $M$ is a deterministic function of the committed cut $C$. In particular, it depends only on finalized lane contents and not on local arrival order, wall clock time, or any executor local state.
\end{prop}

\begin{proof}
Each step of $M$ is defined over data contained in $C$, transaction tips and hashes are fields of finalized transactions, and replica indices are fixed by the configuration. No step references executor local information. Hence any two correct executors holding the same $C$ produce the same output.
\end{proof}

\subsection{Socialised tips}
In Giga the priority fee buys ordering, not a relationship with a particular proposer. Tips are therefore not paid to the proposer whose lane carries the executed copy of a transaction. Instead, all priority fees collected in an epoch are pooled and distributed across the validator set, weighted by stake and by measured liveness.

\begin{definition}[Epoch tip pool]
For epoch $e$, let $T_e$ denote the sum of priority fees charged on transactions executed in blocks finalized during $e$, net of duplicate refunds.
\end{definition}

\begin{definition}[Liveness weight]
For validator $v$ and epoch $e$, let $\lambda^e_v\in[0,1]$ be the fraction of protocol duties performed by $v$ during $e$, where duties are on chain observable events, consensus votes contributed to committed quorum certificates, certified blocks produced in $v$'s lane, and signed divergence attestations. The precise definition and relative weighting of duty classes is a governance parameter.
\end{definition}

\begin{definition}[Tip distribution rule]
A validator $v$ with bonded stake $s_v$ receives
\[
\pi^e_v \;=\; T_e \cdot \frac{s_v\,\lambda^e_v}{\sum_{u} s_u\,\lambda^e_u},
\]
paid at the close of epoch $e$ and shared with its delegators pro rata, as with block rewards.
\end{definition}

Because no lane captures the fees of the transactions it carries, copying a visible high tip transaction into one's own block no longer reassigns its fee, removing the
revenue motive behind the same tick duplicate channel of \cite{landers2025mcpmev}; the residual ordering effects of copying are the concern of the merge rule and of Sedna. Socialisation also makes validator revenue independent of orderflow routing, so there is no fee capture market in steering users toward particular lanes, and the liveness weight converts withholding or lazy participation into a direct revenue penalty. It does not, by itself, eliminate off band side payments for intra lane position; that question, together with the distribution fee and refund flows, belongs
to the full transaction fee mechanism deferred to an upcoming work.

Because transaction validity is only established at execution, admission is rate limited, each validator includes at most one copy of a given transaction per epoch, so a transaction occupies at most $n$ lane slots per epoch and the worst case ordered but unexecuted load per transaction is a protocol constant. Bounding and pricing the analogous load from distinct conflicting transactions admitted against pre-attestation state is upcoming which will remove the need for such accounting.

\section{Sedna: A Practical Private Dissemination Layer}
Sedna is best viewed as a practical alternative to encrypted mempools for a multi-proposer chain. Rather than hiding a single global transaction pool, it replaces whole transaction broadcast with private, addressed, rateless coded dissemination across proposer lanes \cite{ranchal2025sedna}. That framing matters because Sedna is not solving one problem in isolation. It is intended to simultaneously reduce pre-inclusion MEV leakage, improve goodput, and preserve low latency, censorship resistant inclusion. 

\subsection{Protocol overview}
Under Sedna, a sender first commits to a transaction payload and then encodes it into verifiable rateless symbols. The sender does not transmit the full payload to every proposer. Instead, it sends addressed bundles of symbols to selected proposer lanes together with a commitment bound header. Once enough verified symbols have been finalized to cross the decode threshold, executors reconstruct the unique payload and execute it in deterministic order \cite{ranchal2025sedna}. This fits naturally with Sei Giga's architecture, ordering still happens through the existing multi-proposer pipeline, while full payload recovery can occur once the chain has finalized enough information to decode.

\subsection{Why Sedna is a good fit for Sei Giga}
Because only coded fragments are revealed prior to decode, Sedna provides until decode privacy rather than a fully transparent pre-inclusion broadcast model. This substantially reduces the surface for the MCP specific MEV channels discussed above, especially those that depend on other proposers learning the full payload early enough to duplicate, auction, or react to it \cite{ranchal2025sedna,landers2025mcpmev}. Sedna directly attacks the waste created by naive replication. Instead of sending the full transaction to many proposers, the sender distributes smaller addressed bundles. The Sedna paper shows that this rateless approach can approach the information theoretic lower bound for dissemination overhead while requiring no consensus modifications \cite{ranchal2025sedna}. Sedna also preserves the reason to use MCP in the first place. The sender can target enough lanes to achieve a chosen reliability or censorship resistance level without paying the cost of full replication to all of them. In other words, Sedna does not trade away latency to buy privacy, it uses coded dissemination to improve the latency-cost frontier itself \cite{ranchal2025sedna}.

\subsection{Withholding and incentives}
Coded dissemination is not sufficient on its own. A coalition of proposer lanes can receive addressed bundles, withhold them from finalization, and privately accumulate an informational lead while honest executors wait to cross the decode threshold. Our mechanism design follow on to Sedna shows that this problem becomes especially sharp when the decode slack is zero, in which case withholding even a single critical bundle can push decode into the next slot with high probability \cite{marsh2026sednamech}. The proposed answer is to reward the bundles that actually matter for decode. In particular, \textsf{PIVOT-$K$} concentrates a sender funded bounty on the pivotal bundles that trigger decoding, while an adaptive sender ratchet stops contacting lanes whose addressed bundles were not redeemed \cite{marsh2026sednamech}. The practical takeaway for Sei Giga is straightforward, if Sedna is introduced, it should be introduced together with its incentive layer and with parameters that avoid zero slack operation. Under that design, the dissemination layer becomes not just private but economically robust.

\subsection{A pseudo-encrypted mempool}
A threshold encrypted mempool hides transaction payloads until a distributed decryption protocol runs after ordering finality, introducing an additional coordination step and latency on the critical path. Sedna's privacy guarantee, by contrast, expires at the moment the data availability threshold is crossed, a step that Giga's consensus pipeline already requires for liveness. No additional cryptographic coordination round is needed, and no new trust assumption beyond the existing $f{+}1$ availability quorum is introduced. Moreover, an encrypted mempool does not address dissemination cost such that the sender must still replicate the full ciphertext to every proposer for censorship resistance, whereas
Sedna's rateless coding reduces per-lane bandwidth to a fraction of the payload size. In short, Sedna provides comparable pre-execution privacy with lower latency overhead and strictly better goodput.

\section{Block-STM Style Parallel Execution}
\begin{figure}[hbt!]
    \centering
    \includegraphics[width=1\linewidth]{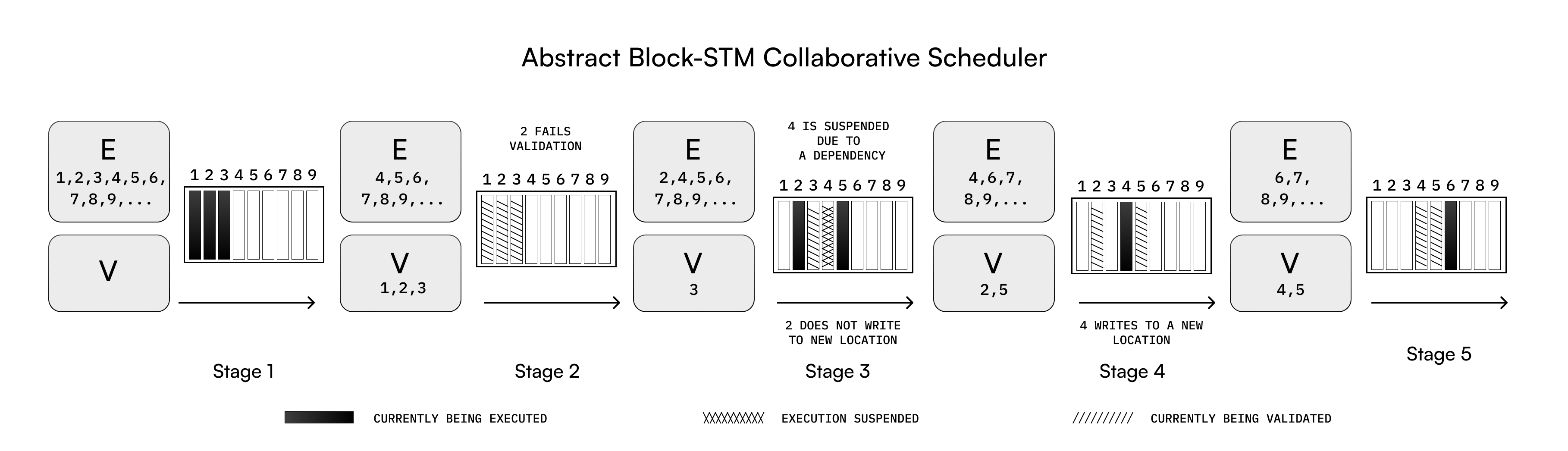}
    \caption{STM}
    \label{fig:stm}
\end{figure}
Let $\{\, t_1, t_2, \ldots, t_n \}$ be the transactions in a finalized block $B$,
arranged in the total order determined by that block. Each transaction $t_i$ reads
some set of addresses, storage slots, or global variables, collectively denoted by
$R_i$, and writes a set of addresses or slots denoted by $W_i$. We say that $t_j$ depends on $t_i$, denoted $t_i \to t_j$, precisely when $j>i$ and
\[
W_i \cap (R_j \cup W_j) \neq \varnothing.
\] 
Intuitively, this captures all cases in which $t_j$ must see the updated value produced by $t_i$. Because the block imposes a total order on transactions, the resulting dependency relation is acyclic.

Sei Giga uses optimistic concurrency control (OCC) to leverage the fact
that most transactions are unlikely to conflict. In OCC, each node begins by
executing all transactions in parallel, distributing them across multiple worker
threads and assuming no conflicts will occur \cite{para}. While executing, each transaction $t_i$ uses a private buffer to store changes to its write set $W_i$, rather than updating the globally visible EVM state. Once $t_i$ finishes, it enters
a validation phase to detect conflicts with any committed transaction $t_k$ where $k < i$. A conflict arises if $t_k$ has written to an address that is in either $R_i$ or $W_i$ after $t_i$ began. If such a conflict is found, $t_i$ is rolled back and re-executed (potentially in a more conservative manner). If no conflict is detected, the changes in $W_i$ are committed to the global state, and $t_i$ is marked complete.

\begin{definition}[Valid parallel schedule]
A parallel execution schedule for block $B=(t_1,\ldots,t_n)$ is \emph{valid} if its committed effect on state, receipts, and gas accounting is identical to the sequential execution of $t_1,\ldots,t_n$ in block order.
\end{definition}

\begin{prop}[Correctness of OCC with rollback]
Assume the executor commits $t_j$ only after every predecessor $t_i$ with $t_i \to t_j$ has committed, and re-executes any transaction whose read or write set is invalidated by an earlier committed transaction. Then the resulting parallel execution is a valid parallel schedule.
\end{prop}

\begin{proof}
A transaction may commit only after every earlier conflicting transaction on which it depends has already committed. Thus committed dependencies are respected in block order. Any optimistic execution that reads stale data or conflicts with an earlier committed write is discarded and recomputed against the updated committed prefix. Therefore the final committed effect is equivalent to sequential execution in block order.
\end{proof}

When contention is low, most transactions commit after one execution attempt. Under sustained contention, the implementation may fall back to sequential execution without changing semantics.

\begin{figure}[hbt!]
    \centering
    \includegraphics[width=1\linewidth]{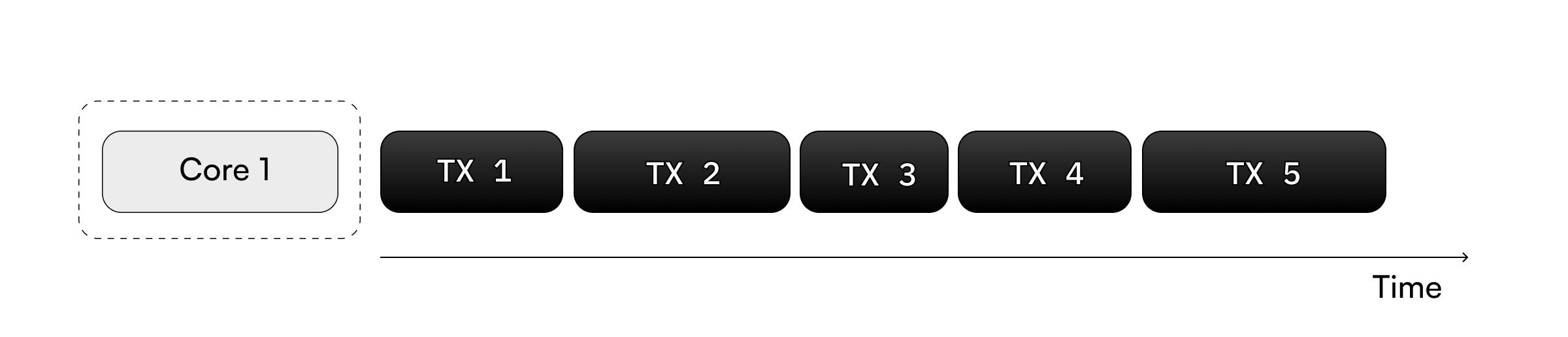}
    \caption{Sequential tx execution}
    \label{fig:seq}
\end{figure}
\begin{figure}[hbt!]
    \centering
    \includegraphics[width=1\linewidth]{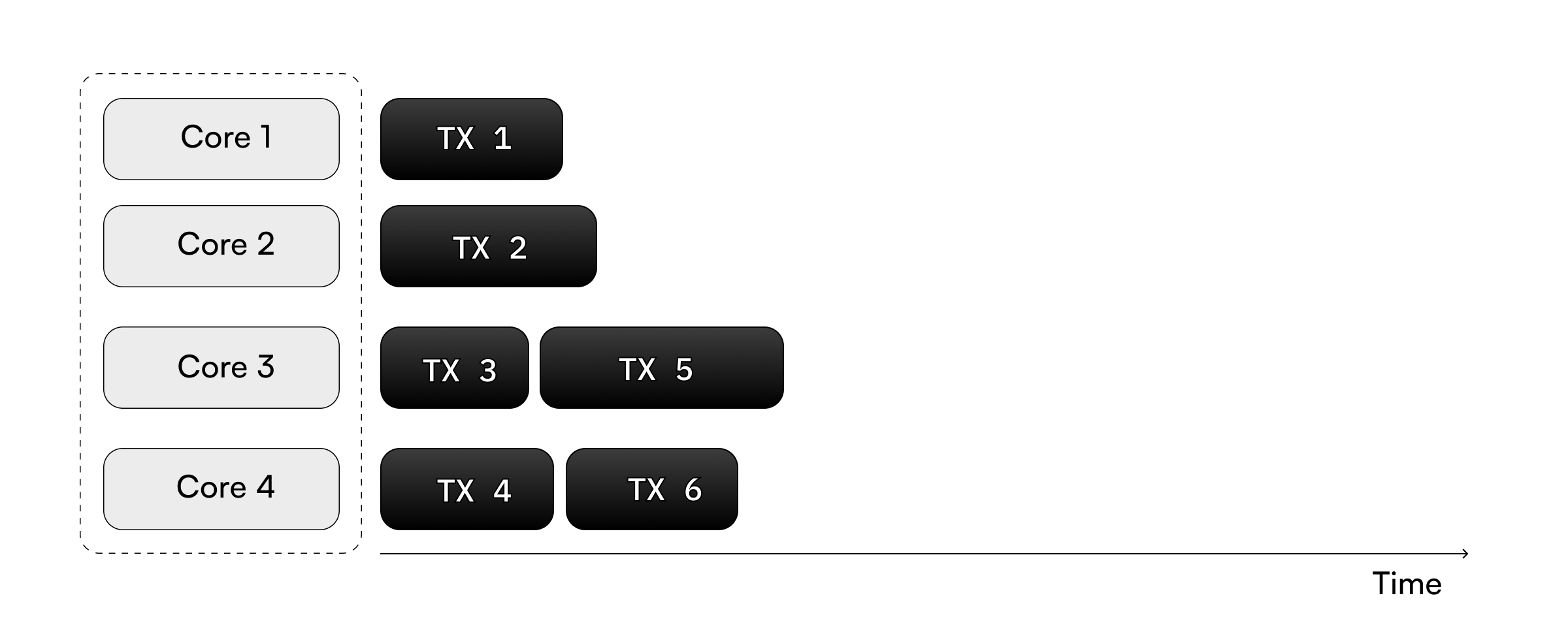}
    \caption{Parallel tx execution}
    \label{fig:occ}
\end{figure}

By allowing for parallel, rather than sequential transaction processing, as shown in figure \ref{fig:seq} and figure \ref{fig:occ} it is trivial to see how execution times can be reduced.
\FloatBarrier
\section{Storage}
Sei Giga adopts a storage strategy that departs from conventional Merkle tree based designs in order to reduce overhead and improve parallelism under extreme throughput. Instead of maintaining a Merkle tree for every update to global state, the chain uses a flat key-value model, where each account, contract storage slot, or globally accessible variable is mapped directly to a corresponding entry in a log structured merge (LSM) tree. Storing \(\langle k, v\rangle\) pairs in this manner is particularly effective for write intensive environments. Sequential updates are efficiently batched and flushed to underlying media.

By eliminating the overhead of traversing or updating multiple tree nodes for each write, a flat key-value design inherently reduces structural churn, streamlines concurrency, and yields more predictable I/O behavior. Because each update targets a single \(\langle k, v\rangle\) pair rather than an entire branch, locking and write contention are simplified, facilitating parallel access across distinct keys without intricate coordination. In addition, storing data in a single level structure reduces pointer-chasing and deep lookups, enabling more efficient batch operations within the underlying LSM engine. Together, these benefits allow Sei Giga to sustain high throughput without the frequent re-hashing and node level bookkeeping overhead that typify Merkle-y approaches.

Sei Giga further reduces storage burdens through a tiered approach. Recent blocks and frequently accessed data reside on high performance local SSDs, ensuring fast lookups and updates. Historical or rarely queried information, by contrast, is migrated to a cold layer that leverages a distributed, columnar database. This cold layer architecture alleviates the need for every validator node to store the complete historical ledger on expensive primary media, it also provides a scalable interface for analytical queries, audit trails, and forensics. As the network processes petabytes of new data each year under a 5 gigagas load, offloading stale state from SSDs to cost effective archival storage becomes central to sustaining practical hardware requirements for validator operators.

Internally, Sei Giga’s implementation leverages an append only write ahead log (WAL) to guard against crashes and data corruption. Updates are written sequentially to the WAL and then applied to the DB, which manages the LSM tree and coordinates compaction routines to maintain balanced input output performance. This mechanism ensures durability, enabling a node to recover from abrupt failures by replaying the WAL and returning the database to a consistent state. Later state consensus attestations are over compact divergence digests for the executed write sets of finalized blocks, preserving a verifiable record of execution agreement without demanding immediate full state re-hashing.

Through this hybrid of a flat key value store, asynchronous execution divergence commitments, tiered storage, and an append only WAL, Sei Giga provides a storage subsystem optimized for continuous high throughput. The approach avoids the overhead typically introduced by Merkle tree or accumulator maintenance on every write, while furnishing a clear growth path for accommodating multiple terabytes of new data annually. This design ensures that user facing nodes can serve recent data efficiently and that validator hardware is not burdened by unnecessary hot path cryptographic bookkeeping.
\subsection{Lattice hash divergence commitments}
After ordering finalizes, honest executors should be able to cheaply attest that they computed the same post block transition and quickly localize divergence if they did not, to do so we use a standard homomorphic hashing approach\cite{cryptoeprint:2019/227}.

\begin{assumption}[Collision resistance on the encoding domain]
Throughout this subsection, we assume that $\LH$ is collision resistant for the deterministic encodings used to compute $d_n$, $d_{n,j}$, and $D_n$.
\end{assumption}

We instantiate $\LH$ as a homomorphic multiset hash \cite{cryptoeprint:2019/227}: let $h:\{0,1\}^*\to\mathbb{Z}_q^{d}$ be a hash modeled as a random oracle, and for a multiset $X$ of records define
\[
\LH(X) \;=\; \sum_{x\in X} h(x) \bmod q .
\]

\begin{assumption}[Multiset collision resistance]
It is infeasible to find distinct multisets $X\neq X'$ of records with $\LH(X)=\LH(X')$. For the parameter regimes of \cite{cryptoeprint:2019/227} this rests on lattice (short integer solution style) assumptions.
\end{assumption}

Let $\Delta_n=\{(k_1,v'_1),\ldots,(k_m,v'_m)\}$ be the write log of finalized block $n$ after intra block last write wins resolution, so each key appears exactly once. Define the record multiset
\[
X_n \;=\;
\{\enc(\mathtt{w},n,k_i,v'_i)\}_{i\le m}
\;\cup\;
\{\enc(\mathtt{r},n,j,R_{n,j})\}_{j}
\;\cup\;
\{\enc(\mathtt{g},n,G_n)\},
\]
where receipt records carry their index $j$ to bind receipt order, and the type tag and height $n$ provide domain separation across record kinds and blocks. The block digest is $d_n=\LH(X_n)$ and the attested commitment is the compact value $D_n=\mathsf{H}(\enc(n,d_n))$ for a standard hash $\mathsf{H}$, the full vector $d_n$ is exchanged only during disputes.

\begin{prop}[Chunk consistency]\label{prop:chunk-consistency}
Partition the key space into ranges $K_1,\ldots,K_s$ and let $d_{n,j}=\LH(\{\enc(\mathtt{w},n,k,v')\in X_n : k\in K_j\})$. Then $\sum_j d_{n,j}$ equals the write component of $d_n$ identically, so chunk digests presented during a dispute are consistent with the attested total by construction rather than by trust.
\end{prop}

\begin{proof}
Each write record lies in exactly one range, and $\LH$ of a disjoint union is the sum of the $\LH$ values of its parts.
\end{proof}

A later block includes a $\frac{2}{3}$ signed attestation to $D_n$. If honest replicas agree on $D_n$, no further action is required. If conflicting commitments appear, replicas compare chunk digests, bisect the disagreement, and replay only the affected range to identify the first divergent write, receipt, or gas accounting error. Because the commitment is over the executed delta rather than the full historical state, the mechanism avoids per-write tree maintenance, pointer chasing, and global re-hashing.
\subsection{Incremental membership proofs (BUDs)}

The divergence digest $D_n$ lets validators attest internal agreement but does
not furnish an external state proof.  A light client or bridge verifier needs a
compact witness for the value of a specific key at a specific height without a
global state trie.  Block Update Digests (BUDs) provide such witnesses with
commitment cost proportional to per-block update volume rather than total state
size; the full construction and proof strategies are described in
\cite{littley2026buds}.

Each state entry is augmented with an 8-byte last-modified block height and a
1-byte serialization version reserved for future schema evolution.

For each key $k_i$ in the canonical write log $\Delta_n$, let $p_i$ denote the
greatest height $p < n$ at which $k_i$ was modified within the BUD-tracked
history, setting $p_i = -1$ when no such height exists.  Define the block
update digest $U_n$ to be the Merkle root of the lexicographically sorted
leaves
\[
  (k_i,\; v'_i,\; n,\; p_i),
\]
after resolving intra-block duplicates in favour of the last write in block
order.  $U_n$ is computed from the same write log as $D_n$ and attested on the
same delayed $\frac{2}{3}$-signed schedule; an implementation may attest a
joint digest encoding both.

A BUD proof is a Merkle inclusion path from a leaf $(k,v,n,p)$ to an attested
root $U_n$.  It certifies that $k$ held value $v$ immediately after block $n$
and records the prior modification height $p$.  If two valid BUD proofs for the
same key $k$ exist at heights $a < b$ and the leaf at $b$ records
previous-modification height $a$, then $k$ is unmodified throughout $(a,b)$
and the value at $a$ holds for every height in $[a,b)$.  A \emph{touch
transaction} rewrites only the last-modified metadata of a key, forcing a fresh
BUD leaf at the current height and providing an immediate proof anchor for
stale keys.

To bridge large gaps without a touch, Giga maintains a fixed aligned
exponential hierarchy of aggregated digests called SuperBUDs.  Fix a branching
base $e \ge 2$ and a maximum level $L_{\max}$.  A level-$\ell$ SuperBUD
$S_{\ell,j}$ is the Merkle-trie root of the partial map
$k \mapsto M_{\ell,j}(k)$, where $M_{\ell,j}(k)$ is the greatest block height
at which $k$ was modified within the aligned window
\[
  W_{\ell,j} \;=\; \bigl[\,H_{\mathrm{BUD}} + j\,e^\ell,\;\;
                         H_{\mathrm{BUD}} + (j{+}1)\,e^\ell - 1\,\bigr].
\]
Adjacent same-level SuperBUDs merge into the next level by a single $O(n)$
lexicographic traversal retaining only the more recent modification height per
key.  The parameters $e$ and $L_{\max}$ are governance-configurable and define
the guaranteed proof window $\eta = e^{L_{\max}}$.  Claims whose nearest BUD
anchor is older than $\eta$ may be served by archive nodes via SuperBUD
concatenation but fall outside the protocol guarantee.

Deleted keys are represented by tombstones preserving the modification chain;
tombstones older than $\eta$ are garbage-collected.  An exclusion proof anchors
on a tombstone at height $d \le h$ and uses BUD or SuperBUD witnesses to show
no modification in $(d,h]$.  Keys with no post-activation history require a
touch transaction to create the initial witness.

At activation height $H_{\mathrm{BUD}}$, a one-time synthetic write log
bootstraps all extant keys into the BUD subsystem with predecessor sentinel
$-1$.  No proof is defined for $h < H_{\mathrm{BUD}}$.  The SuperBUD hierarchy
reaches steady state after $\eta$ blocks; during ramp-up, proofs may
concatenate lower-level SuperBUDs at the cost of increased proof size.  The BUD
and SuperBUD Merkle trees use a classical hash; their short-lived proof window
makes classical collision resistance sufficient, and the hash function falls
under the same post-quantum migration schedule described in Section~8.

\section{Post-Quantum Migration}
Sei Giga inherits the classical key assumptions of the EVM account model. If a large scale quantum attacker can break the classical signature scheme protecting an account, that attacker can seize control of the account once its public key is exposed. A full redesign of address derivation and wallet interfaces is possible, but it is too large a dependency for an initial migration. A more credible short term plan is to preserve existing accounts by binding a post-quantum verification key to each account before a governance-set cutoff while classical signatures are still trusted \cite{nistfips204,nistpqcfaq,nistpqcproject,nistir8547}.

\subsection{Design goal}
Our short term goal is not to make the entire account model elegant. It is to survive a Q-day event without forcing a chain wide account reset. The migration therefore optimizes for continuity of existing balances, cheap validator side lookup, and minimal hot path complexity. It deliberately sacrifices unrestricted creation of new EOAs after the cutoff.

\subsection{Pre-Q-day key registration}
Let $H_Q$ denote the activation height after which classical ECDSA style authorization is no longer accepted for externally owned accounts. Before $H_Q$, any existing account $a$ may submit a registration transaction
\[
\Pi[a] = (\mathsf{alg}_a, pk^{PQ}_a, \nu_a, h_a)
\]
signed by its current classical key. Here $\mathsf{alg}_a$ identifies the post-quantum scheme, $pk^{PQ}_a$ is the registered verification key, $\nu_a$ is a migration nonce, and $h_a$ is an optional activation height satisfying $h_a \leq H_Q$. The registration transaction may also require a proof-of-possession under $pk^{PQ}_a$ so that an address cannot be bound to an unusable or third party key.

During a transition window $[H_{\mathrm{reg}},H_Q)$ the chain may accept either classical signatures or a dual signature format in which the same payload is signed under both the classical and post-quantum keys. This preserves backwards compatibility for wallets while classical cryptography remains trusted.

\subsection{Post-Q-day verification}
After height $H_Q$, authorization is table driven. For a transaction claiming sender address $a$, the execution client performs a direct lookup in the registered key table. If $\Pi[a]=\bot$, the transaction is invalid. Otherwise the client verifies
\[
\mathsf{Verify}_{PQ}\bigl(pk^{PQ}_a, m, \sigma^{PQ}\bigr)=1.
\]
There is no post-Q-day public key recovery path and no creation of fresh EOAs from new classical keys. In other words, after $H_Q$ an account can spend only if it was bound to a post-quantum verification key before the cutoff. This gives the execution client the desired hash table like verification path, resolve sender address, fetch registered key, verify signature, continue. This design is intentionally conservative as it keeps existing accounts live, avoids ambiguous post-cutoff key discovery, and prevents an attacker from introducing new classical accounts after the point at which classical signatures are no longer trusted. The trade off is that genuinely new user onboarding must occur through pre-registration, contract wallets, or a later native post-quantum account format. Giga intends, in the short term, to use ML-DSA \cite{nistfips204}. It is standardized now, NIST explicitly accommodates dual signatures during transition, and the current PQC project guidance is that the first PQC standards can be put into use now \cite{nistpqcfaq,nistpqcproject}.



\section{Future work}
A full tokenomics overview of Giga is not included in this version of the whitepaper and will be covered in a future update. Said work will include the full TFM to be used by Giga, building on top of several WIP papers. As noted in the introduction to this paper the pricing model for Giga will in its simplest form consist for 3 distinct components, the gas fee, the priority fee for ordering, and the distribution fee allowing the network to price duplicates sent for censorship resistance even when they're not executed. Although the Sei token will continue to be used for fees, staking, and rewards, the full economic landscape including slashing mechanisms and reward functions are deferred for now. This work only describes the short term quantum upgrade path which is known to not be performant enough for Giga, the proposal described in this work is intended as a temporary solution that adds little overhead to the current protocol but can be relied upon in an emergency. Work is ongoing on full Giga scale PQC alternatives which will be covered in subsequent works. Sedna work is also ongoing with several other works in progress which will influence the final version of Sedna that will be used in Giga, along with consensus upgrades dealing with the single leader in Autobahn. The BUD parameters $e$ and $L_{\max}$ will be calibrated from empirical proof-generation benchmarks once the execution client is deployed at scale.

\printbibliography
\end{document}